\documentclass[journal,twocolumn]{IEEEtran}

\usepackage{enumerate}
\usepackage{amsmath,amsthm}
\usepackage{mathtools}
\usepackage{algorithm,algorithmic}
\usepackage{float}
\usepackage{hyperref}
\usepackage{color}
\usepackage{makeidx}
\usepackage{bbm}
\usepackage{graphicx}
\usepackage{lipsum}
\usepackage{soul}
\usepackage{tabularx}
\usepackage{dsfont}
\usepackage[table,xcdraw]{xcolor}

\usepackage{amsfonts}
\usepackage{times}
\usepackage{graphicx}
\usepackage{latexsym}
\usepackage{dsfont}
\usepackage{amssymb}
\usepackage{amsmath}
\usepackage{cite}
\usepackage{verbatim}
\usepackage{subfigure}

\def\bb0{{\mathbb{0}}}

\def\bb{{\mathbf{b}}}

\def\bw{{\mathbf{w}}}

\def\b0{{\mathbf{0}}}

\def\sf0{{\mathsf{0}}}

\usepackage{epstopdf}
\usepackage{balance}
\newcommand{\sref}[1]{{Section}~\ref{#1}}
\newcommand{\fref}[1]{{Fig.}~\ref{#1}}
\newcommand{\aref}[1]{{Algorithm}~\ref{#1}}

\DeclareMathOperator*{\argmax}{arg\,max}

\newcommand{\subto}{\operatorname{s.t.}}
\newcommand{\argmin}{\operatornamewithlimits{arg\min}}

\begin{document}
\title{Reinforcement Learning for Beam Pattern Design in Millimeter Wave and Massive MIMO Systems}
\author{Yu Zhang, Muhammad Alrabeiah, and Ahmed Alkhateeb \thanks{Yu Zhang, Muhammad Alrabeiah and Ahmed Alkhateeb are with Arizona State University (Email: y.zhang, malrabei, alkhateeb@asu.edu). This work is supported by the National Science Foundation under Grant No. 1923676.}}
\maketitle

\begin{abstract}
Employing large antenna arrays is a key characteristic of  millimeter wave (mmWave) and terahertz communication systems. However, due to the adoption of  fully analog or hybrid analog/digital architectures, as well as  non-ideal hardware or arbitrary/unknown array geometries, the accurate channel state information becomes hard to acquire. This impedes the design of beamforming/combining vectors that are crucial to fully exploit the potential of large-scale antenna arrays in providing sufficient receive signal power.  In this paper, we develop a novel framework that leverages deep reinforcement learning (DRL) and a Wolpertinger-variant architecture and learns how to iteratively optimize the beam pattern (shape) for serving one or a small set of users relying only on the receive power measurements and without requiring any explicit channel knowledge.   The proposed model accounts for key hardware constraints such as the phase-only, constant-modulus, and quantized-angle constraints. Further, the proposed framework can efficiently optimize the beam patterns for systems with non-ideal hardware and for arrays with unknown or arbitrary array geometries. Simulation results show that the developed solution is capable of finding near optimal beam patterns based only on the receive power measurements.
\end{abstract}

\section{Introduction} \label{intro}

Leveraging the large bandwidth available at millimeter wave (mmWave) frequency bands requires the deployment of large antenna arrays. However, because of the high cost and power consumption of the mixed-circuit components, mmWave systems normally rely either fully or partially on analog beamforming, where transmitters/receivers employ networks of quantized phase shifters \cite{Alkhateeb2014MIMO,Alkhateeb2014}. This makes the basic MIMO signal processing functions, such as channel estimation, challenging as the channels are seen only through the RF lens. As a result, classical beamforming/combining design approaches, e.g. \cite{Lo1999,Love2003}, may not be feasible because of the unavailability of the channels as well as the new constraints of the design problem. Besides, the hardware is possibly not ideal due to the use of inexpensive and low-precision radio components. In this case, the performance of the commonly used beams (such as the ones in classical beamsteering codebooks) degrades drastically for their unawareness of the environment and hardware/array geometry.

\textbf{Prior Work:}
Designing efficient beamforming and combining  is essential for realizing the  potential of MIMO communications, and it has been an important research topic in the literature of MIMO signal processing \cite{Lo1999,Love2003, Li2017, ElAyach2014, Alkhateeb2014MIMO}.  For MIMO systems with no hardware constraints, i.e., with fully-digital processing and no constraints on the RF hardware, maximum ratio transmission and combining maximize the achievable SNR with single-stream transmission/reception  \cite{Lo1999}. To realize these solutions, however, the MIMO system should be able to control the magnitude and phase of the signal at each antennas. When only the phase can be controlled, equal-gain transmission solutions have been developed to maximize the SNR or diversity gains \cite{Love2003}. This is particularly interesting for mmWave and terahertz systems where the beamforming/precoding processing is fully or partially done in the RF domain using analog phase shifters \cite{Alkhateeb2014MIMO}. In these systems, however, the phase shifters can normally take only quantized phase shift values. This associates the search over the large space of quantized phase shift values with high complexity (e.g., for a 32-element antenna array with 2-bit phase shifters,  there are $4^{32}$ possible beamforming vectors) \cite{ Li2017, ElAyach2014, Alkhateeb2014MIMO}. Further, in analog beamforming architectures,  the channel is seen through the RF lens, which makes it hard to acquire at the baseband, especially for systems with arbitrary or unknown array geometries. To address these challenges, this paper designs a reinforcement learning based approach to efficiently learn the analog beamforming patterns that adapt to the surrounding environment and the adopted hardware/array geometry  without requiring explicit channel knowledge.

\textbf{Contribution:}
In this paper, we propose a deep reinforcement learning based framework that can learn how to optimize the beam pattern for serving a single user or a set of users with similar channels. \textbf{The developed framework relies only on receive power measurements and does not require any channel knowledge.} This framework adapts the beam pattern based on the surrounding environment and learns how to compensate for the hardware impairments. This is done by utilizing a novel Wolpertinger architecture \cite{Dulacarnold2015} which is designed to efficiently explore the large discrete action space. The proposed model accounts for key hardware constraints such as the phase-only, constant-modulus, and quantized-angle constraints \cite{Alkhateeb2014MIMO}. This is realized by defining the state directly as the phases of the analog phase shifters and the action as the change of phases within the quantized phase set. Simulation results show that the proposed solution is capable of finding the near optimal beam pattern and achieving a beamforming/combining gain comparable to that of equal gain combining.

\section{System and Channel Models} \label{sec:System}

In this section, we introduce in detail our adopted system and channel models. We also describe how the model considers arbitrary array geometries with possible hardware impairments.

\subsection{System Model}

We consider the system model where a mmWave massive MIMO base station (BS) with $M$ antennas is communicating with a single-antenna user. Further, given the high cost and power consumption of mixed-signal components, we consider a practical system where the BS has only one radio frequency (RF) chain and employs analog-only beamforming/combining using a network of $r$-bit quantized phase shifters. Therefore, the beamforming/combining vector can be written as
\begin{equation}\label{Analog}
  {\bf w} = \frac{1}{\sqrt{M}}\left[ e^{j\theta_1}, e^{j\theta_2}, \dots, e^{j\theta_M} \right]^T,
\end{equation}
where each phase shift $\theta_m$ is selected from a finite set $\boldsymbol{\Theta}$ with $2^r$ possible discrete values drawn uniformly from $(-\pi, \pi]$.
In the uplink transmission, if a user $u$ transmits a symbol $x\in\mathbb{C}$ to the base station, where the transmitted symbol satisfies the average power constraint $\mathbb{E}\left[|x|^2\right]=P_x$, the received signal at the base station after combining can be expressed as
\begin{equation}\label{sys}
  y_u = {\mathbf w}^H{\mathbf h}_ux + {\mathbf w}^H{\mathbf n},
\end{equation}
where ${\mathbf h}_u\in\mathbb{C}^{M\times 1}$ is the uplink channel vector between the user $u$ and the base station antennas and ${\mathbf n}\sim\mathcal{N}_\mathbb{C}\left(0, \sigma_n^2{\bf I}\right)$ is the receive noise vector at the base station.

\subsection{Channel Model} \label{subsec:channel}

We adopt a general geometric channel model for ${\mathbf h}_u$. Assume that the signal propagation between the user $u$ and the base station consists of $L$ paths. Each path $\ell$ has a complex gain $\alpha_\ell$ and an angle of arrival $\phi_\ell$. Then, the channel vector can be written as
\begin{equation}\label{channel}
  {\mathbf h}_u = \sum\limits_{\ell=1}^{L}\alpha_\ell{\mathbf a}(\phi_\ell),
\end{equation}
where ${\bf a}(\phi_\ell)$ is the array response vector of the base station. The definition of ${\bf a}(\phi_\ell)$ depends on the array geometry and hardware impairments. Next, we discuss that in more detail.

\begin{figure*}[t]
	\centering
    \includegraphics[width=.95\textwidth]{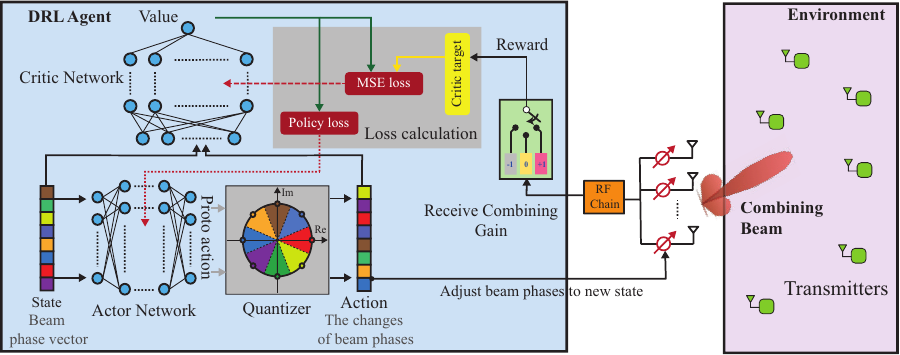}
	\caption{The proposed beam pattern design framework with deep reinforcement learning. The schematic shows the agent architecture, and the way it interacts with the environment.}
	\label{BP-Alg}
\end{figure*}

\subsection{Hardware Impairments Model} \label{subsec:impairments}

Most of the prior work on mmWave signal processing has assumed uniform antenna arrays with perfect calibration and ideal hardware \cite{Hur2013,Wang2009,Alkhateeb2014,Alkhateeb2014MIMO}. In this paper, we consider a more general antenna array model that accounts for arbitrary geometry and hardware impairments, and target learning beam pattern that mitigates the influence of those unknown factors.
While the beam pattern learning solution that we develop in this paper is general for various kinds of array geometries and hardware impairments, we evaluate the proposed solution in \sref{sec:Results} with respect to two main characteristics of interest, namely non-uniform spacing and phase mismatch between the antenna elements. For linear arrays, the array response vector can be modeled to capture these characteristics as follows
\begin{multline}\label{ARV-cor}
  {{\bf a}}(\phi_\ell) = \left[ e^{j\left(kd_1\cos(\phi_\ell) + \Delta\theta_1\right)}, e^{j\left(kd_2\cos(\phi_\ell) +\Delta\theta_2\right)},\dots, \right. \\
  \left. e^{j\left(kd_M\cos(\phi_\ell) + \Delta\theta_M\right)} \right]^T,
\end{multline}
where $d_m$ is the position of the $m$-th antenna, and $\Delta\theta_m$ is the additional phase shift incurred at the $m$-th antenna (to model the phase mismatch). Without loss of generality, we assume that $d_m$ and $\Delta\theta_m$ are fixed yet unknown random realizations drawn from the distributions $\mathcal{N}\left((m-1)d, \sigma_d^2\right)$ and $\mathcal{N}\left(0, \sigma_p^2\right)$ respectively, where $d$ is the ideal antenna spacing, $\sigma_d$ and $\sigma_p$ model the standard deviations of the random antenna position and phase mismatch. Besides, we impose an additional constraint $d_1<d_2<\cdots<d_M$ to make sure the generated antenna positions physically meaningful.

\section{Problem Definition} \label{sec:Prob}

In this paper, we investigate the beam pattern design problem for mmWave and massive MIMO system with unknown array geometry and hardware impairment. Given the system and channel models described in \sref{sec:System}, the SNR after combining for user $u$ can be written as
\begin{equation}\label{single_snr}
  \mathsf{SNR}_u = \frac{\left|{\bf w}^H{\bf h}_u\right|^2}{\left\|{\bf w}\right\|^2} \rho = \left|{\bf w}^H{\bf h}_u\right|^2 \rho,
\end{equation}
where $\left\|{\bf w}\right\|^2=1$ is implicitly used and $\rho=\frac{P_x}{\sigma_n^2}$. Besides, we define the beamforming/combining gain of adopting $\bw$ as a transmit/receive beamformer for user $u$ as
\begin{equation}\label{single_bfgain}
  g_u = \left| {\bf w}^H{\bf h}_u \right|^2.
\end{equation}
It can be seen that maximizing \eqref{single_bfgain} is equivalent to maximizing the SNR in \eqref{single_snr}.
Therefore, the objective of this paper is to design (learn) the beamformer ${\bf w}$ that maximizes the beamforming/combining gain given by \eqref{single_bfgain} for a single user or a set of users with similar channels. Therefore, the beam pattern learning problem can be formulated as
\begin{align}\label{Prob-1}
 {\bf w}_{\mathsf{opt}} = \argmax\limits_{{\bf w}} & \hspace{2pt}  \frac{1}{|\boldsymbol{\mathcal{H}}|}\sum_{{\bf h}_u\in\boldsymbol{\mathcal{H}}} \left| {\bf w}^H{\bf h}_u \right|^2, \\
 \subto  \hspace{2pt} &  w_{m} = \frac{1}{\sqrt{M}}e^{j\theta_{m}}, ~ \forall m=1, ..., M, \label{cons-1} \\
 & \theta_{m}\in\boldsymbol{\Theta}, ~ \forall m=1, ..., M, \label{cons-2}
\end{align}
where $w_m$ is the $m$-th element of the beamforming vector and $\boldsymbol{\mathcal{H}}$ is the channel set that is supposed to contain a single channel or multiple similar channels.
It is worth mentioning that the constraint in \eqref{cons-1} is imposed to uphold the adopted analog-only system model, and the constraint in \eqref{cons-2} is to respect the quantized phase-shifters hardware constraint.

Due to the unknown array geometry as well as possible hardware impairments, the accurate channel state information is generally hard to acquire. This means that all the channels ${\bf h}_u\in\boldsymbol{\mathcal{H}}$ in the objective function are possibly unknown. Instead, the base station may only have access to the beamforming/combining gain $g_u$, or equivalently the Received Signal Strength Indicator (RSSI). Therefore, problem \eqref{Prob-1} is hard to solve in a general sense for the unknown parameters in the objective function as well as the non-convex constraint \eqref{cons-1} and the discrete constraint \eqref{cons-2}. Given that \textbf{this problem is essentially a search problem in a dauntingly huge yet finite and discrete space,} we consider leveraging the powerful exploration capability of deep reinforcement learning to efficiently search over the space to find the optimal or near-optimal solution.

\section{Beam Pattern Learning} \label{sec:BPL}

In this section, we present our proposed DRL-based algorithm for addressing the beam pattern design problem \eqref{Prob-1}. It is worth mentioning that when viewing the problem from a reinforcement learning perspective, it features a \textbf{finite yet very high dimensional} action space. This makes the traditional learning frameworks (such as deep Q-learning, deep deterministic policy gradient, etc.) hard to apply. Therefore, we adopt a novel architecture called Wolpertinger to enable the efficient search in a large discrete action space, the details of which can be found at \cite{Dulacarnold2015}.

\subsubsection{Reinforcement Learning Setup}

To solve the problem with reinforcement learning, we first specify the corresponding building blocks of the learning algorithm as follows:
\begin{itemize}
  \item \textbf{State:} We define the state ${\bf s}_t$ as a vector that consists of the phases of all the phase shifters at the $t$-th iteration, that is, ${\bf s}_t=\left[\theta_1, \theta_2, \dots, \theta_M\right]^T$. This phase vector can be converted to the actual beamforming vector by applying \eqref{Analog}. Since all the phases in ${\bf s}_t$ are selected from $\boldsymbol\Theta$, and all the phase values in $\boldsymbol\Theta$ are within $(-\pi, \pi]$, \eqref{Analog} essentially defines a bijective mapping from the phase vector to the beamforming vector. Therefore, for simplicity, we will use the term ``beamforming vector'' to refer to both this phase vector and the actual beamforming vector (the conversion is given by \eqref{Analog}), according to the context.
  \item \textbf{Action:} We define the action ${\bf a}_t$ as the element-wise changes to all the phases in ${\bf s}_t$. Since the phases can only take values in $\boldsymbol\Theta$, a change of a phase means that the phase shifter selects a value from $\boldsymbol\Theta$. Therefore, the action is directly specified as the next state, i.e. ${\bf s}_{t+1}={\bf a}_t$. 
  \item \textbf{Reward:} We define a ternary reward mechanism, i.e. the reward $r_t$ takes values from $\{+1, 0, -1\}$. We compare the beamforming gain achieved by the current beamforming vector, denoted by $g_t$, with two values: (i) an adaptive threshold $\beta_t$, and (ii) the previous beamforming gain $g_{t-1}$. The reward is computed using the following rule
      \begin{itemize}
        \item $g_t > \beta_t$, $r_t=+1$;
        \item $g_t \le \beta_t$ and $g_t > g_{t-1}$, $r_t=0$;
        \item $g_t \le \beta_t$ and $g_t \le g_{t-1}$, $r_t=-1$.
      \end{itemize}
\end{itemize}

It is important to note that the adopted adaptive threshold mechanism does not rely on any prior knowledge of the channel distribution. The threshold value starts from zero and whenever the BS tries a new beam and the resulting beamforming gain surpasses the current threshold, the system updates the threshold by the value of this new beamforming gain. Besides, since the update of threshold also marks a successful detection of a new beam that achieves the best beamforming gain so far, the BS also records this beamforming vector. As can be seen in the reward definition, in order to calculate the reward, the system always tracks two quantities, which are the previous beamforming gain and the best beamforming gain achieved so far (i.e. the threshold).

\begin{algorithm}[ht]
	\caption{DRL Based Beam Pattern Learning}
	\label{alg1}
	\begin{algorithmic}[1]
    \STATE Initialize actor network $\mu({\bf s}|\theta^\mu)$ and critic network $Q({\bf s}, {\bf a}|\theta^Q)$ with random weights $\theta^\mu$ and $\theta^Q$
    \STATE Initialize target networks $\mu^\prime$ and $Q^\prime$ with the weights of actor and critic networks' $\theta^{\mu^\prime}\leftarrow\theta^\mu$ and $\theta^{Q^\prime}\leftarrow\theta^Q$
    \STATE Initialize the replay memory $\mathcal{D}$, minibatch size $B$
    \STATE Initialize adaptive threshold $\beta=0$ and the previous average beamforming gain $g_1=0$
    \STATE Initialize a random process $\mathcal{N}$ for action exploration
    \STATE Initialize a random phase vector as the initial state ${\bf s}_1$
    \FOR{$t=1$ to $T$}
        \STATE Receive a predicted action from actor network with exploration noise $\widehat{\bf a}_t = \mu({\bf s}_t|\theta^\mu) + \mathcal{N}_t$
        \STATE Quantize the predicted action to a valid beamforming vector ${\bf a}_t$ according to \eqref{quant}
        \STATE Execute action ${\bf a}_t$, observe reward $r_t$ and update state to ${\bf s}_{t+1}={\bf a}_t$
        \STATE Update the threshold $\beta$ and previous gain $g_t$
        \STATE Store the transition $({\bf s}_t, {\bf a}_t, r_t, {\bf s}_{t+1})$ in $\mathcal{D}$
        \STATE Sample a random mini batch of $B$ transitions $({\bf s}_b, {\bf a}_b, r_b, {\bf s}_{b+1})$ from $\mathcal{D}$
        \STATE Calculate target $y_b=r_b+\gamma Q^\prime({\bf s}_{b+1}, \mu^\prime({\bf s}_{b+1}|\theta^{\mu^\prime})|\theta^{Q^\prime})$
        \STATE Update the critic network by minimizing the mean squared loss $L=\frac{1}{B}\sum_{b}(y_b-Q({\bf s}_b, {\bf a}_b|\theta^Q))^2$
        \STATE Update the actor network using the sampled policy gradient given by

        $-\frac{1}{B}\sum_{b=1}^{B} \nabla_{{\bf a}}Q({\bf s}, {\bf a})|_{{\bf s}={\bf s}_b, {\bf a}=\mu({\bf s}_b|\theta^\mu)}\nabla_{\theta^\mu}\mu({\bf s}|\theta^\mu)|_{{\bf s}={\bf s}_b}$
        \STATE Update the target networks every $C$ iterations
    \ENDFOR
	\end{algorithmic}
\end{algorithm}

\subsubsection{Environment Interaction} \label{LF}

As mentioned in Sections \ref{intro} and \ref{sec:Prob}, due to the possible hardware impairments, accurate channel state information is generally unavailable. Therefore, the base station can only resort to the beamforming/combining gain to adjust its beam pattern in order to achieve a better performance. Upon forming a new beam $\tilde{\bw}$, the base station uses this beam to receive the pilots transmitted from every user. Then, it averages all the beamforming gains
\begin{equation}\label{avg_bf_eval}
  \bar{g} = \frac{1}{|\boldsymbol{\mathcal{H}}|}\sum_{{\bf h}_u\in\boldsymbol{\mathcal{H}}} \left| \tilde{\bw}^H{\bf h}_u \right|^2,
\end{equation}
where $\boldsymbol{\mathcal{H}}$ represents the targeted user channel set. Recall that \eqref{avg_bf_eval} is the same as evaluating the objective function of \eqref{Prob-1} with the current beamforming vector $\tilde{\bw}$. Depending on whether or not the new average beamforming gain surpasses the previous one as well as the current threshold, the base station gets either reward or penalty, based on which it can judge the ``quality'' of the current beam and decide how to move.

\subsubsection{Exploration} \label{subsub:Explore}

The exploration happens after the actor network predicts the action $\widehat{\bf a}_{t+1}$ based on the current state (beam) ${\bf s}_{t}$. Upon obtaining the predicted action, an additive noise is added element-wisely to $\widehat{\bf a}_{t+1}$ for the purpose of exploration, which is a customary way in the context of reinforcement learning with continuous action spaces \cite{Sutton2018,Timothy2015}. In our problem, we use temporally correlated noise samples generated by an Ornstein-Uhlenbeck process \cite{Uhlenbeck1930}, which is also used in \cite{Dulacarnold2015}. It is worth mentioning that a proper configuration of the noise generation parameters has significant impact on the learning process. Normally, the extent of exploration (noise power) is set to be a decreasing function with respect to the iteration number, which is commonly known as exploration-exploitation tradeoff \cite{Sutton2018}.
Furthermore, the exact configuration of noise power should relate to specific applications. In our problem, for example, the noise is directly added to the predicted phases. Thus, at the very beginning, the noise should be strong enough to perturb the predicted phase to any other phases in $\boldsymbol\Theta$. By contrast, when the learning process approaches to the termination (the learned beam already performs well), the noise power should be decreased to a smaller level that is only capable of perturbing the predicted phase to its adjacent phases in $\boldsymbol\Theta$.

\subsubsection{Quantization} \label{subsub:Quantize}

The predicted beam (with exploration noise added) should be quantized in order to be a valid new beam that can be implemented by the discrete phase-shifters. Therefore, each quantized phase in the new vector can be calculated as
\begin{equation}\label{quant}
  [{\bf s}_{t+1}]_m = \argmin_{\theta\in\boldsymbol{\Theta}}\left|\theta-[\widehat{{\bf s}}_{t+1}]_m\right|, \forall m=1, 2, \dots, M,
\end{equation}
which is essentially a nearest neighbor lookup (i.e. a KNN classifier with $k=1$).

\subsubsection{Forward Computation and Backward Update} \label{subsub:FB}

The current state ${\bf s}_{t}$ and the new state ${\bf s}_{t+1}$ (recall that we directly set ${\bf s}_{t+1}={\bf a}_t$) are then fed into the critic network to compute the Q value, based on which the targets of both actor and critic networks are calculated. This completes a forward pass. Following that, a backward update is performed to the parameters of the actor and critic networks. A pseudo code of the algorithm can be found in \aref{alg1}.

\begin{figure}[t]
  \includegraphics[width=.95\linewidth]{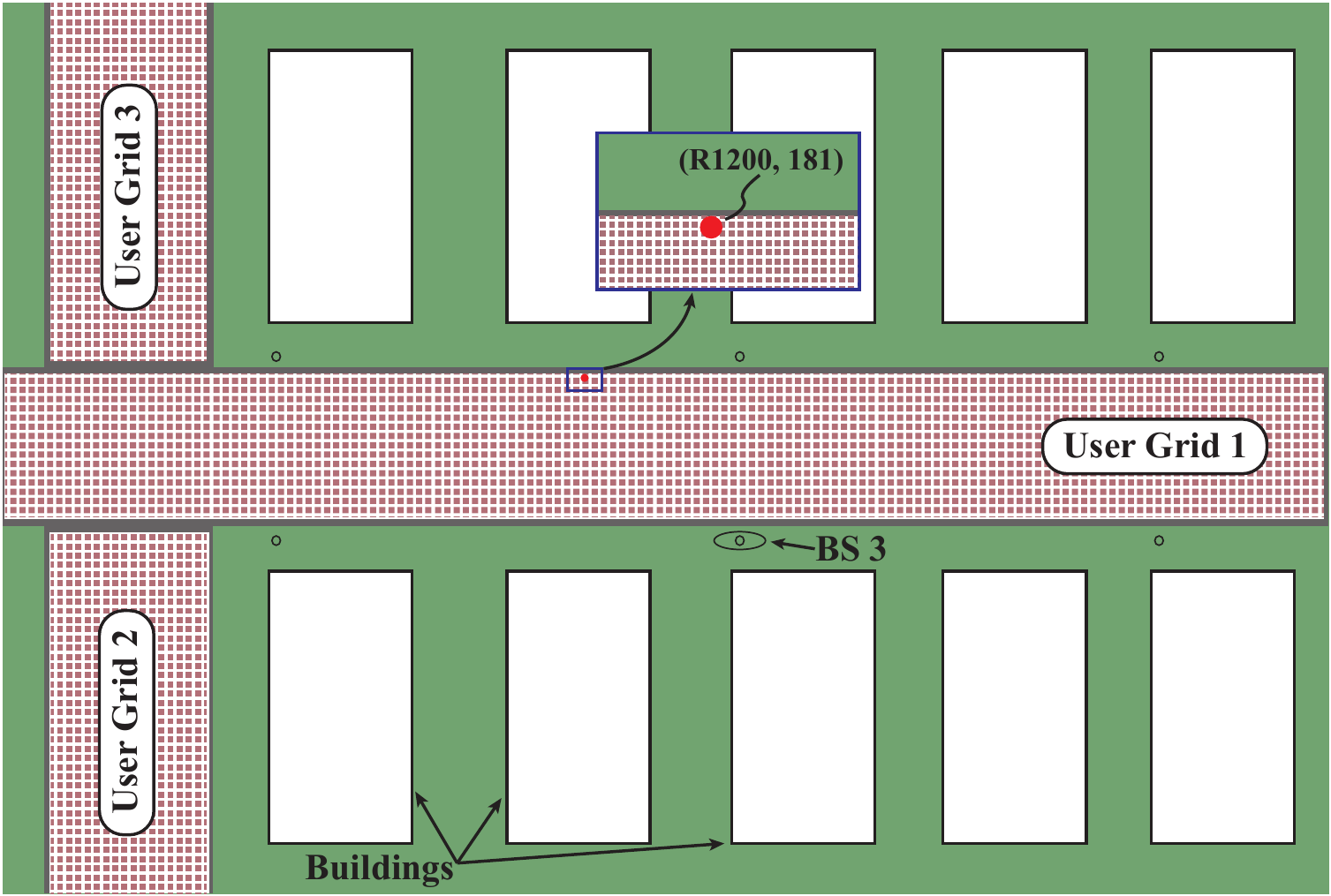}
  \caption{The top view of the considered communication scenario.}
  \label{fig:sce_los}
\end{figure}

\section{Simulation Results} \label{sec:Results}

In this section, we evaluate the performance of the proposed solution. We first describe the adopted scenario and dataset used in our simulations and then discuss the results.

\subsection{Scenario and Dataset}

In our simulations, we consider the outdoor scenario `O1\_60' which is offered by the DeepMIMO dataset \cite{DeepMIMO} and is generated based on the accurate 3D ray-tracing simulator Wireless InSite \cite{Remcom}. This scenario comprises two streets and one intersection with three uniform x-y user grids, as shown in \fref{fig:sce_los}. To generate the channels from the users to the base station, we adopt the following DeepMIMO parameters: (1) Scenario name: O1\_60, (2) Active BSs: 3, (3) Active users: Row 1200 to 1200, (4) Number of BS antennas in (x, y, z): (1, 32, 1), (5) System bandwidth: 1 GHz, (6) Number of OFDM sub-carriers: 1 (single-carrier), (7) Number of multipaths: 5.
From the generated dataset, we further select the user at row 1200 and column 181 in the scenario. The locations of both the selected user and the base station are marked in \fref{fig:sce_los}.

\begin{figure*}[t]
	\centering
    \includegraphics[width=.95\textwidth]{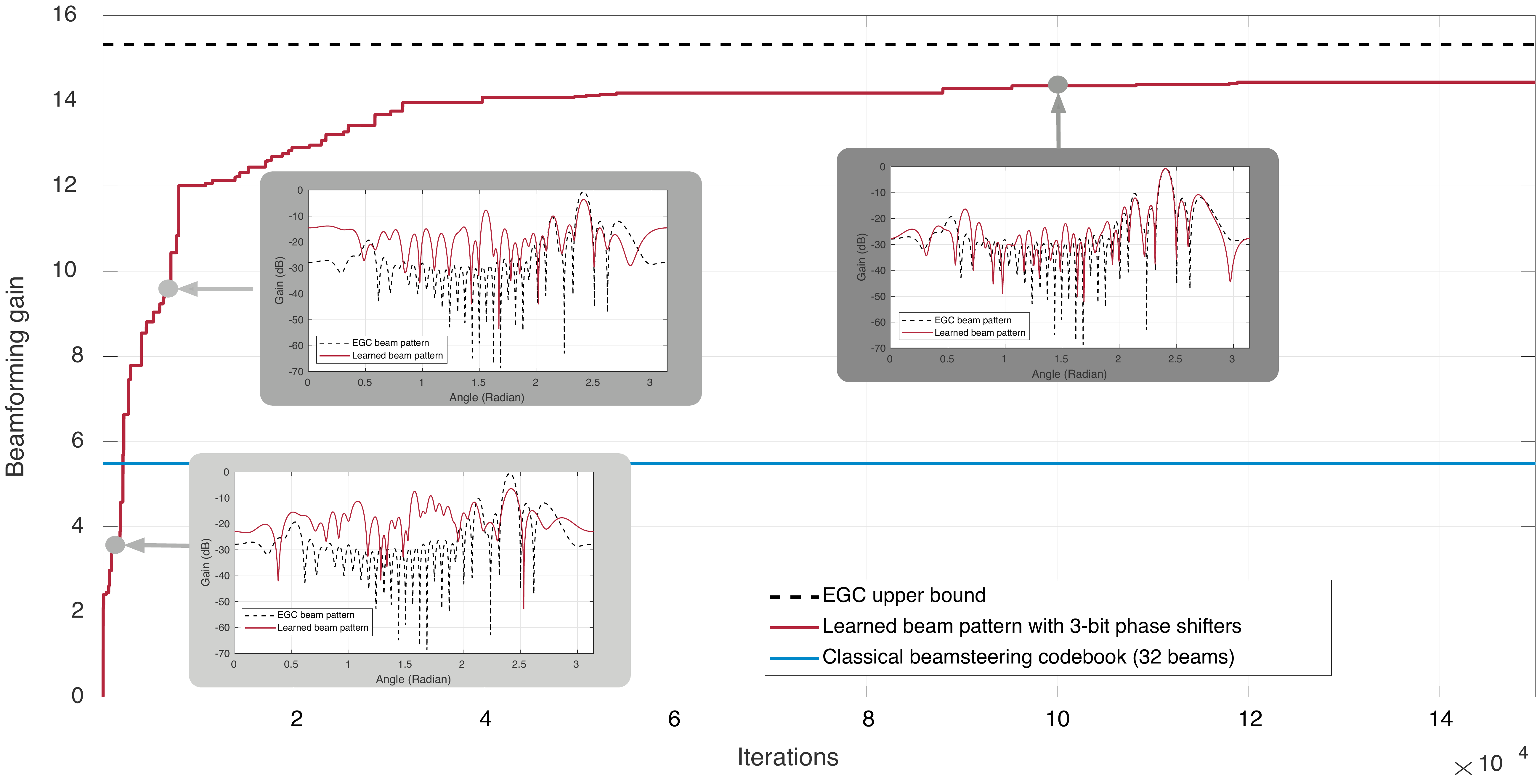}
	\caption{The beam pattern learning results for a single user with LOS connection to the base station. The base station employs a perfect uniform linear array with 32 antennas and 3-bit phase shifters. In this figure, we show the learning process and the beam patterns learned at three different stages during the iterations. The learned beam patterns are plotted using solid red line, and the equal gain combining/beamforming vector is plotted using dashed black line.}
	\label{1b-clean}
\end{figure*}

\subsection{Performance Evaluation}

We first evaluate our proposed DRL-based beam pattern learning solution on learning a single beam that serves a single user with LOS connection to the base station.
In \fref{1b-clean}, we compare the performance of the learned single beam with a 32-beam classical beamsteering codebook. As it is commonly known, classical beamsteering codebook normally performs very well in LOS scenario. However, our proposed method achieves higher beamforming gain than the best beam in the classical beamsteering codebook, with negligible iterations. More interestingly, with less than $4\times10^4$ iterations, the proposed solution can reach more than $90\%$ of the EGC upper bound. It is worth mentioning that the EGC upper bound can only be reached when the user's channel is known and unquantized phase shifters are deployed. By contrast, our proposed solution can finally achieve almost $95\%$ of the EGC upper bound with 3-bit phase shifters and without any channel information.

We also plot the learned beam patterns at three different stages (iteration 1000, 5000, and 100000) during the learning process, which helps understand how the beam pattern evolves over time. As shown in \fref{1b-clean}, at iteration 1000, the learned beam pattern has very strong side lobes, weakening the main lobe gain to a great extent. At iteration 5000, the gain of the main lobe becomes stronger. However, there are still multiple side lobes with relatively high gains. Finally, at iteration 100000, it can be seen that the main lobe has quite strong gain compared to the other side lobes, having at least 10 dB gain over the second strongest side lobe. And most of the side lobes are below $-20$ dB. Besides, the learned beam pattern captures the EGC beam pattern very well, which explains the good performance it achieves. The slight mismatching is mainly caused by the use of quantized phase shifters. With 3-bit resolution, each phase shifter can only realize 8 different values of phase shifts drawn uniformly from $(-\pi, \pi]$.

The proposed beam pattern learning solution is also evaluated on a system where hardware impairments exist (with the same user considered above). This is a more realistic and interesting scenario, for mmWave systems are susceptible to hardware mismatches like antenna spacing mismatch and phase mismatch. The wavelength in mmWave bands is so small that even slight mismatching can lead to a drastic degradation of the performance. This for sure calls for an intelligent design process that is capable of adapting the beam pattern to the hardware, mitigating the loss caused by hardware mismatches. The simulation results confirm that our proposed solution is competent to learn such optimized beam pattern for a system with hardware impairments.

\begin{figure}[t]
	\centering
	\subfigure[]{ \includegraphics[width=0.43\linewidth]{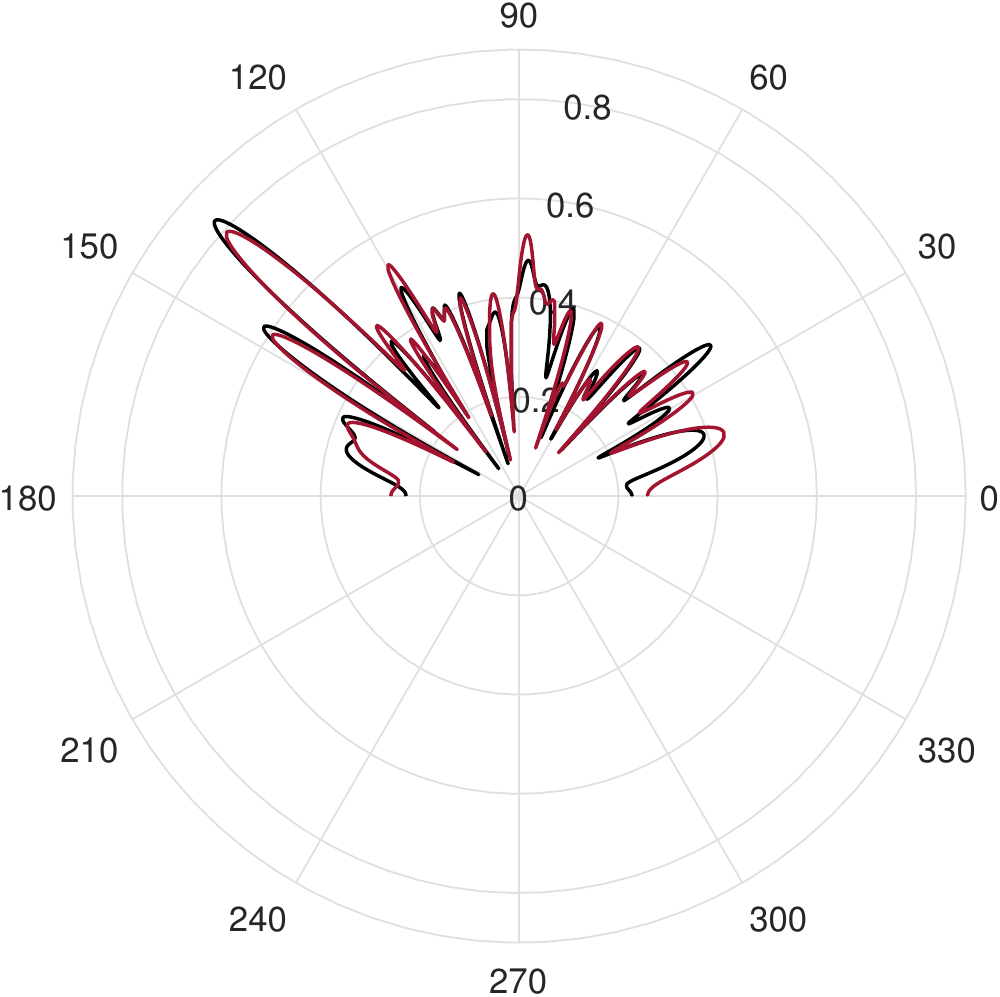} }
	\subfigure[]{ \includegraphics[width=0.5\linewidth]{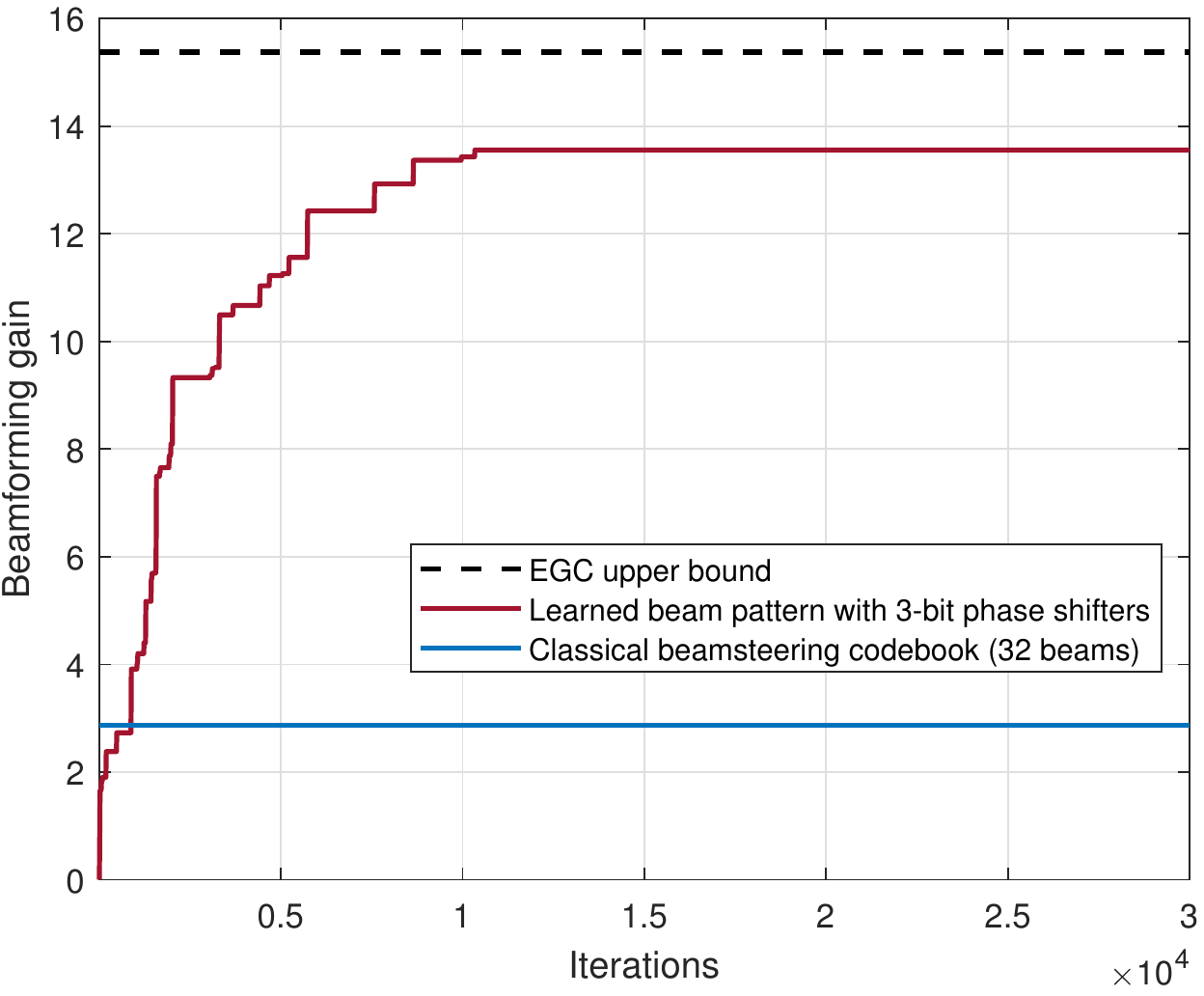} }
	\caption{The beam pattern learned for a single user with LOS connection to the base station. The base station employs a uniform linear array with 32 antennas and 3-bit phase shifters, where hardware impairments exist. The standard deviation of the antenna position is $0.1\lambda$ and the standard deviation of the phase mismatches is $0.32\pi$. (a) shows the beam patterns for the equal gain combining/beamforming vector (black) and the learned beam (red). A transformation of $\sqrt[4]{\cdot}$ is used to better show the finer structure of the beams. (b) shows the learning process.}
	\label{1b-corrupted}
\end{figure}

\fref{1b-corrupted} (a) shows the beam patterns for both equal gain combining/beamforming vector (plotted in black) and the learned beam (plotted in red). At the first glance, the learned beam appears distorted and has multiple low-gain lobes. However, the performance of such beam is excellent. This can be explained by comparing the beam patterns of the learned beam and the equal gain combining/beamforming vector. \textbf{As can be seen from the learned beam patterns, our proposed solution intelligently approximates the optimal beam, where all the dominant lobes are well captured.} By contrast, the classical beamsteering codebook fails when the hardware is not perfect, as depicted in \fref{1b-corrupted} (b). This is because the distorted array pattern incurred by the hardware impairment makes the pointed classical beamsteering codebook beams only able to capture a small portion of the transmitted power, which further results in an inferior beamforming/combining gain. The learned beam shown in \fref{1b-corrupted} (a) is capable of achieving more than $90\%$ of the EGC upper bound with approximately only $10^4$ iterations, as shown in \fref{1b-corrupted} (b). This is especially interesting given that the proposed solution does not rely on any channel state information. As it is known, the channel estimation in this case relies first on a full calibration of the hardware, which is a hard and expensive process.

\section{Conclusions and Discussions} \label{sec:Conclusions}

In this paper, we developed a DRL-based approach to learn the optimized beam pattern for a single user or a group of users with similar channels relying only on the receive power measurements and without any channel knowledge. This approach relaxes the coherence/synchronization requirements and is important for fully-analog or hybrid analog/digital architectures that are commonly adopted by mmWave/terahertz communication systems. The proposed learning framework respects key hardware constraints such as the phase-only, constant-modulus, and quantized-angle constraints. Simulation results show that the proposed solution is capable of finding the near optimal beam pattern which achieves a beamforming/combining gain comparable to that of equal gain combining without any explicit channel knowledge.

\balance

\end{document}